\documentclass[10pt,
							 twocolumn,
							 superscriptaddress,
							 english,
							 prl,
							 showpacs,
							 floatfix,
							 aps
							]{revtex4-1}
\usepackage[utf8]{inputenc}
\usepackage{amsmath}
\usepackage{amssymb}
\usepackage{graphicx}
\usepackage{xspace}

\makeatletter
\newcommand{\Figref}[1]{Figure~\ref{#1}}
\newcommand{\mos}{MoS$_2$}
\newcommand{\Pc}{H$_2$Pc}
\newcommand{\HPc}{HPc}
\newcommand{\didv}{d$I$/d$V$}

\usepackage{babel}

\begin{document}

\title{$\pi$-Radical Formation by Pyrrolic H Abstraction of Phthalocyanine Molecules on Molybdenum Disulfide}

\author{ Ga\"el Reecht}
\email{greecht@zedat.fu-berlin.de}

\author{ Nils Krane}
\author{ Christian Lotze}
\author{ Katharina J. Franke}
\affiliation{Fachbereich Physik, Freie Universit\"at Berlin, Arnimallee 14, 14195 Berlin, Germany.}


\date{\today}

\begin{abstract}
For a molecular radical to be stable, the environment needs to be inert. Furthermore, an unpaired electron is less likely to react chemically, when it is placed in an extended orbital. Here, we use the tip of a scanning tunneling microscope to abstract one of the pyrrolic hydrogen atoms from phthalocyanine (\Pc) deposited on a single layer of molybdenum disulfide (\mos) on Au(111). We show the successful dissociation reaction by current-induced three-level fluctuations reflecting the inequivalent positions of the remaining H atom in the pyrrole center. Tunneling spectroscopy reveals two narrow resonances inside the semiconducting energy gap of \mos\ with their spatial extent resembling the highest occupied molecular orbital (HOMO) of \Pc. By comparison to simple density functional calculations of the isolated molecule, we show that these correspond to a single occupation of the Coulomb-split highest molecular orbital of HPc. We conclude that the dangling $\sigma$ bond after N--H bond cleavage is filled by an electron from the delocalized HOMO. The extended nature of the HOMO together with the inert nature of the \mos\ layer favor the stabilization of this radical state.

\end{abstract}

\maketitle 


Molecular radicals hold one or more unpaired electrons in their valence orbitals. These molecules have attracted great interest for applications in organic optoelectronic \cite{Hattori2014,Peng2015} and spintronic \cite{Mas-Torrent2011, Ratera2012} devices. Unfortunately, the open-shell nature renders these molecules highly reactive. Hence, the class of stable molecular radicals is limited. Most promising candidates for stable radicals are molecules with the unpaired electron being hosted in an extended orbital \cite{Mas-Torrent2011,Shimizu2018}. Such molecules can be obtained, \textit{e.g.}, by dehydrogenation or dehalogenation, where the dissociation reaction involves the highest occupied molecular orbital (HOMO) \cite{Zhao2005, Repp2006a, Pavlicek2015, Pavlicek2017, Pavlicek2017a, Pavlicek2017b}. Prime examples for H abstraction are porphyrin and phthalocyanine molecules, where one of the pyrrolic H atoms can be detached by a voltage pulse from the tip of a scanning tunneling microscope (STM) \cite{Auwaerter2012, Smykalla2014}. However, when the molecules are placed on a metal substrate the expected radical state is not preserved. Intramolecular relaxations and charge transfer with the substrate lead to a closed-shell configuration \cite{Sperl2011, Pavlicek2015, Pham2016}. To suppress the interaction of radical molecules with the substrate, thin insulating layers of NaCl have been employed successfully for several radical species  \cite{Repp2006a,Pavlicek2015, Pavlicek2017, Pavlicek2017a, Pavlicek2017b}, but a radical state of H-abstracted Pc has not been reported to date.

Here, we show that a monolayer of \mos\ on Au(111) is an ideal system to study the H-abstraction reaction by STM. We prove the successful dehydrogenation reaction by three-level fluctuations of the tunneling current, reflecting the switching of the position of the single H atom in the pyrrolic center \cite{Auwaerter2012}. By mapping the spatial distribution of the singly occupied molecular orbital (SOMO) and singly unoccupied molecular orbital (SUMO), we find that the cleavage of an N--H bond does not lead to a localized dangling bond, but to an extended $\pi$ radical hosted by the former HOMO. We propose that intramolecular charge transfer thus stabilizes the radical in an extended orbital, similar to intramolecular charge transfer when both H atoms are detached from the molecule \cite{Neel2016}.

\begin{figure}[t]
\begin{center}
\includegraphics[width=1\columnwidth]{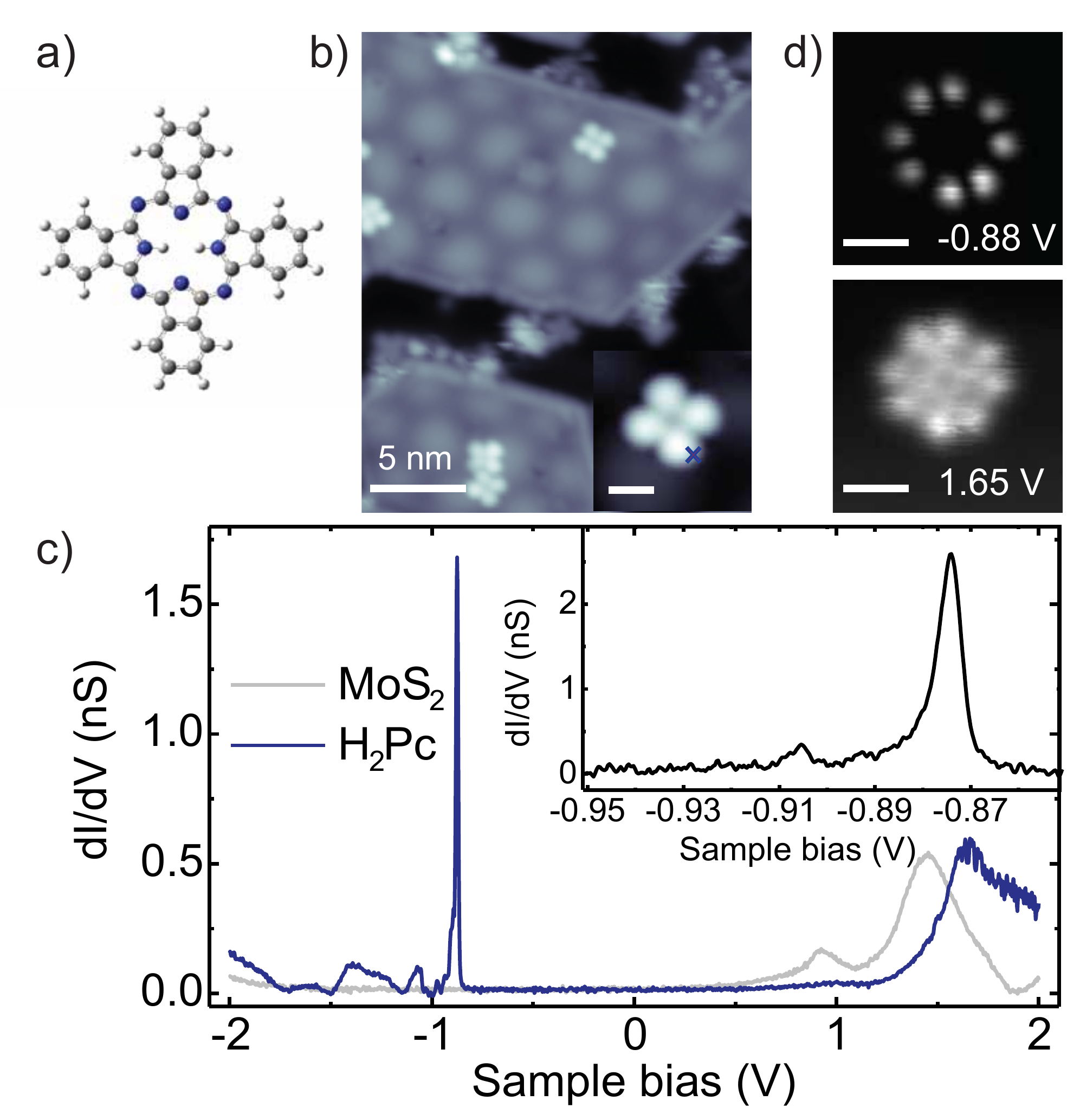}
\end{center}
\caption{a) Structure model of the \Pc\ molecule. (b) 20\,x\,25 nm$^2$ STM topography of \mos\ islands on Au(111) with isolated \Pc\ molecules (V=0.5\,V, I=10\,pA). Close-up view of a single \Pc\ molecule is shown in the inset (scale bar is 1\,nm). c) \didv\ spectra recorded on the bare \mos\ (gray line) and on the \Pc\ molecule (blue line) at the position indicated by the cross in b) (Set point: 2\,V, 300\,pA, $V_\text{mod}$=5\,mV). Inset: High-resolution spectrum around the energy of the sharp resonance at negative bias (set point: -1.2\,V, 50\,pA, $V_\text{mod}$=1\,mV). d) Experimental constant-height \didv\ maps of the the resonance energies of \Pc\ (scale bars are 1\,nm).}
\label{fig1}
\end{figure}

STM images of monolayer islands of \mos\ on Au(111) show a characteristic moir\'e pattern due to the lattice mismatch of the S-terminating lattice and the Au substrate (\Figref{fig1}b) \cite{Gronborg2015, Krane2016}. Deposition of a low density of \Pc\ molecules at a sample temperature of 120\,K leads to a random distribution of isolated molecules, both, on the bare Au(111) surface as well as the \mos\ islands. In both locations the molecules appear with a fourfold clover shape (at sufficiently low bias voltage) (inset in \Figref{fig1}b). The shape essentially corresponds to the topographic structure of the \Pc\ molecules, because the low bias voltage lies within the HOMO-LUMO gap of the molecule (see below) \cite{Sperl2011, Neel2016, Kugel2017, Imai-Imada2018}.

Differential conductance spectra on \mos\ reflect the semiconducting band gap around the Fermi level (gray spectrum in \Figref{fig1}c) \cite{Sorensen2014, Krane2016}. Spectra recorded on one of the isoindole units of the \Pc\ molecules on \mos\ exhibit a sharp conductance peak at around -0.88\,V and a broad peak at 1.65\,V. The narrow resonance lies within the gap of the \mos\ monolayer. Its width of  $\sim$ 7\,meV (inset of \Figref{fig1}c) signifies a long lifetime ($\sim$ 100\,fs) of the transiently charged molecular state. The resonance is followed by a small vibronic side peak at -0.905\,V.  Such a narrow linewidth and the observation of vibronic levels within the excited state are ascribed to an efficient electronic decoupling from the metal substrate combined with a small electron-phonon coupling strength in the underlying layer (\mos), preventing fast relaxation of the excitation \cite{Krane2018}.
Differential-conductance maps at the bias voltage of the sharp resonance show eight lobes -- two lobes on each isoindole moiety (\Figref{fig1}d). This shape has been identified as the HOMO of the \Pc\ molecule on other decoupling layers \cite{Neel2016,Imai-Imada2018} and is in agreement with gas-phase DFT calculations (see below). The broad conductance peak at 1.65\,V lies outside the energy gap of \mos. Hence, the excitation of the negative ion resonance is not protected by the \mos\ layer. A conductance map recorded at the respective energy shows an eight-lobed intensity on the isoindole units as well as intensity in the center of the macrocycle (\Figref{fig1}d). This shape corresponds to the overlap of the quasi-degenerate LUMO and LUMO+1 of the \Pc\ (see below)\cite{Imai-Imada2018}.
Note that the noise in the images is an inherent molecular property. As we will show below and explain in more detail in Supporting Information, we can associate it to the tautomerization reaction of the two hydrogen atoms within the macrocycle.

\begin{figure}[ht]
\begin{center}
\includegraphics[width=1\columnwidth]{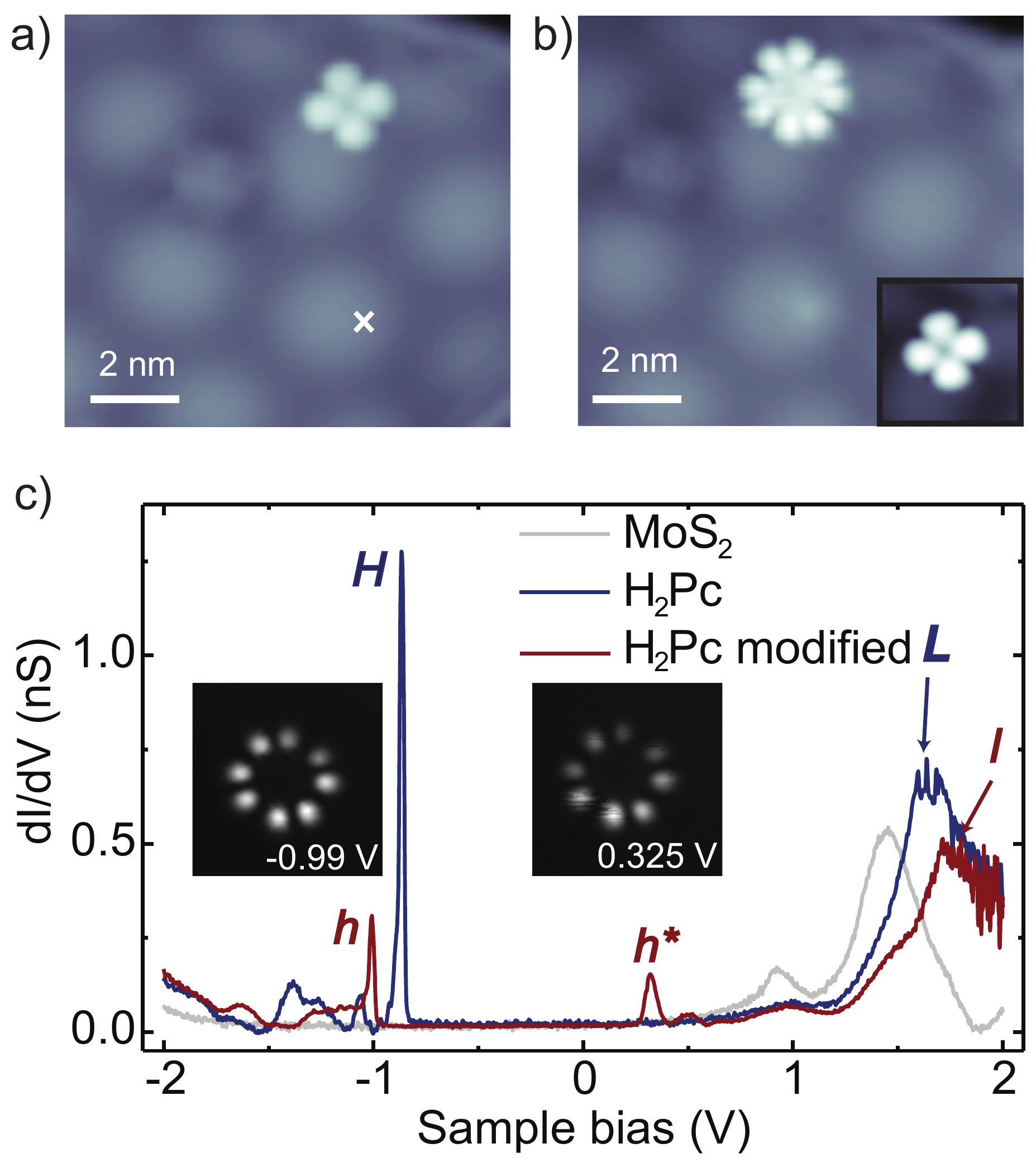}
\end{center}
\caption{a-b) STM topographies before (a) and after (b) a voltage pulse of 3.5\,V was applied at the position indicated by the white cross (V=0.5\,V, I=20\,pA). Inset: Topography of the same molecule with V=0.1\,V.  c) \didv\ spectra on the bare \mos\ (gray line), and on the molecule shown  in the topographies in a) (blue line) and in b) (red line) (Set point: 2\,V, 300\,pA and $V_\text{mod}$=10\,mV)). Inset: Constant-height \didv\ maps recorded at the energy of the states labeled  as  $h$ and $h^*$ in the red spectrum.}
\label{fig2}
\end{figure}

Having characterized the electronic energy levels of the \Pc\ molecules, we now turn to their intentional modification under the influence of a bias-voltage pulse.  \Figref{fig2}a and b show the STM topographies recorded with a sample bias of 0.5\,V, before and after a voltage pulse of 3.5\,V has been applied at the position marked by the white cross. After the voltage pulse, the molecule does not appear with the four-lobe structure but instead exhibits eight lobes at a bias voltage of 0.5\,V. The clover shape is recovered when imaging with a lower bias voltage of 0.1\,V, suggesting that the macrocycle is intact (inset \Figref{fig2}b).

To gain further insights into the modification, we record a \didv\ spectrum on this molecule. The spectrum differs drastically from the initial \Pc\ molecule. Most prominent, a new resonance within the former HOMO--LUMO gap appears at 0.3\,V. This state lies close to the onset of the conduction band of \mos. It is therefore slightly broader than the former states deep inside the bandgap of \mos\ but still very narrow compared to the states outside the gap. Additionally, the other two resonances, \textit{i.e.}, both the narrow resonance at negative bias voltage and the broad resonance at large positive bias voltage, appear shifted away from the Fermi level. Conductance maps at the energies of both narrow resonances reveal the typical eight-lobe structure of the HOMO of the \Pc\ molecule. The same appearance of both orbital distributions at positive and negative bias voltage suggest that they both derive from the former HOMO, which is now singly occupied \cite{Repp2006a}. We therefore label them as $h$ and $h^*$, respectively. From the spectra, we now understand that resonance $h^*$ confers the eight-lobe appearance to the molecule in STM images of \Figref{fig2}b. Note that this tip induced modification of the molecule is easily reproducible, with the similar observations than we describe above (See additional examples in Supporting Information). 

\begin{figure}[ht]
\begin{center}
\includegraphics[width=1\columnwidth]{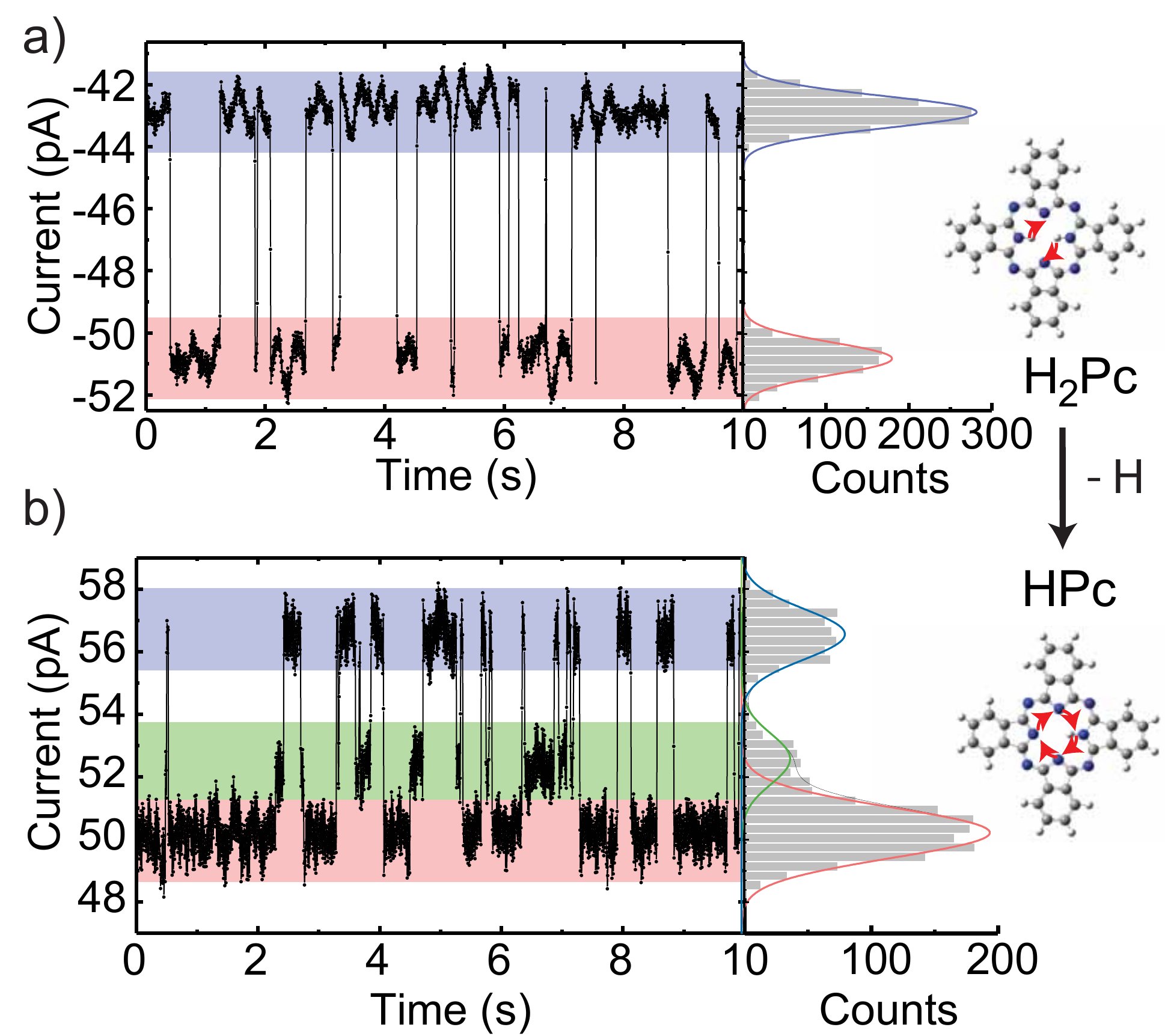}
\end{center}
\caption{Constant-height current traces recorded with the STM tip positioned over one lobe of the phthalocyanine molecule before (a) and after modification (b) by a voltage pulse. Set points before switching off the feedback are 50\,pA and -1.1\,V and 0.2\,V for \Pc\ and HPc, respectively. A histogram of the current distribution is plotted next to each trace, and fitted with two and three Gaussian peaks, respectively. The telegraph noise between distinct levels, indicated by the colored boxes and the Gaussian function for the distributions, stems from tautomerization reactions, as indicated by the schemes on the right. }
\label{fig3}
\end{figure}

To elucidate the origin of the modification of the electronic structure of the phthalocyanine we perform the following experiments. First, we position the STM tip over one lobe of the non-modified \Pc\ and switch off the feedback regulation (constant-height condition). We then record a time trace of the tunneling current. Such a trace shows a bi-stable telegraph noise (\Figref{fig3}a) revealing a two-level fluctuation of the molecule. Similar fluctuations have been observed in free-base phthalocyanine and porphyrin systems and ascribed to the tautomerization reaction within the \Pc\ core \cite{Liljeroth2007, Auwaerter2012, kumagai2014, Kugel2016, Kugel2017}. The tautomerization is a collective switching of the two hydrogen atoms between the two trans configurations inside the pyrrole macrocycle (upper scheme in \Figref{fig3}). Second, we perform the equivalent procedure on the modified molecule. These time traces show a three-level switching (\Figref{fig3}b). An increase in the number of available states can only be explained by the abstraction of one H atom \cite{Auwaerter2012,Kugel2017}. The remaining H atom can then switch between four bonding sites and, hence, four distinct states. However, when the tip is located above one of the four isoindole units, two states are symmetric with respect to the tip position. Thus, they yield the same conductance level, thereby reducing the measurement to a three-level fluctuation instead of four-level fluctuation. We also note that small displacements of the tip from the symmetry points should lead to a distinction as four-level switching. However, the difference in the conductance level is expected to be very small and therefore well below our noise level.
From the distinct tautomerization traces we conclude that the modified molecule occurred upon pyrrolic hydrogen removal under the influence of the voltage pulse.

\begin{figure}[ht]
\begin{center}
\includegraphics[width=1\columnwidth]{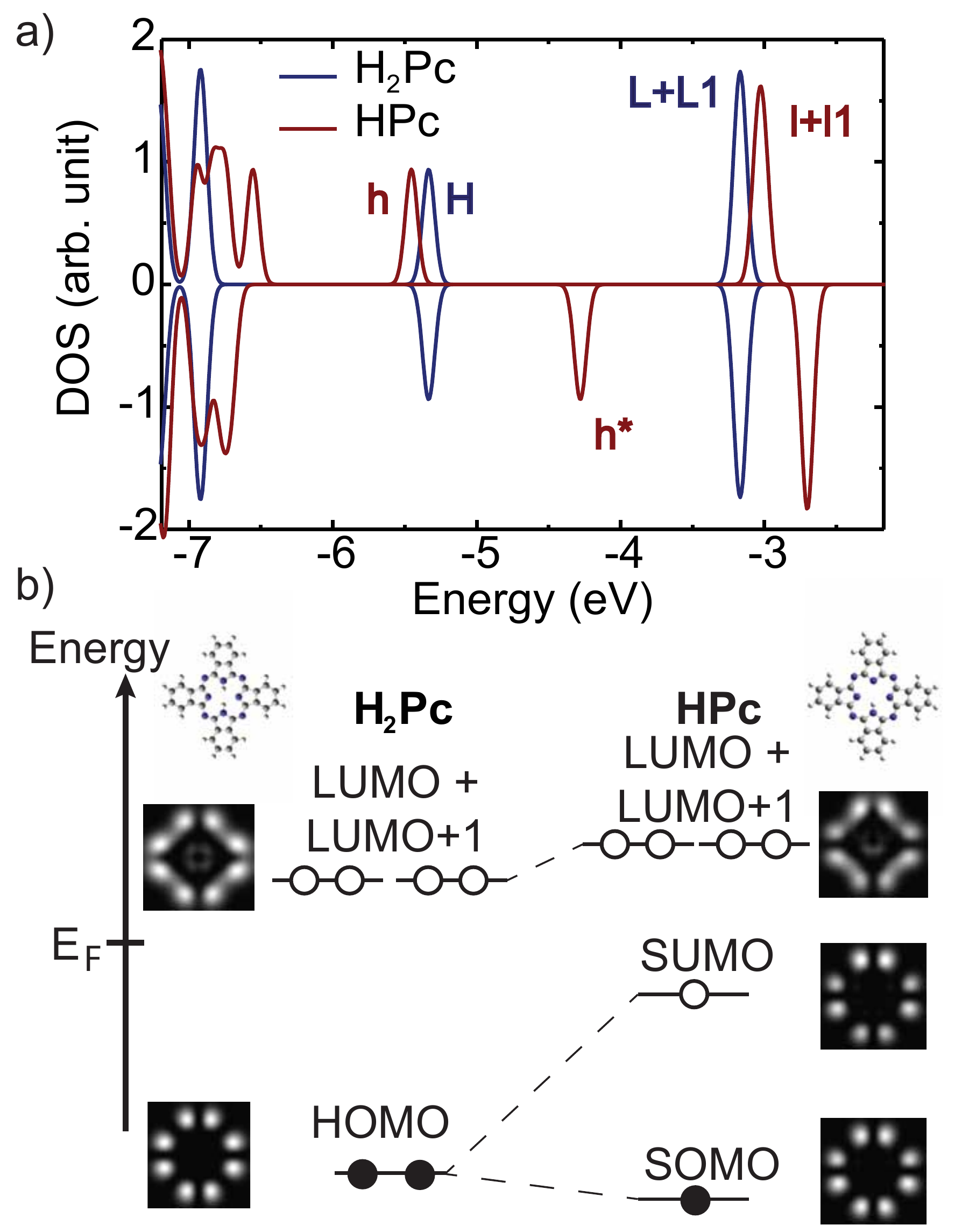}
\end{center}
\caption{a) Spin-resolved calculated partial density of states (PDOS) of \Pc\ (blue line) and HPc (red line) in gas phase. b) Sketch of the energy level diagram  for \Pc\ and HPc with the simulated constant-height conductance maps. The removal of one H atom lifts the spin-degeneracy of the HOMO, leading to one singly-occupied and one singly-unoccupied molecular orbital (SOMO and SUMO). These can be associated to the $h$ and $h*$ resonances in the STS data on HPc. The calculated maps show a larger intensity on the isoindoline unit opposite to the position of the H atom in the pyrrole center. This reduced symmetry is not found in the experimental maps in \Figref{fig2}c because the switching rate is faster than the image acquisition time.}
\label{fig4}
\end{figure}

Despite the clear signature of H abstraction, the molecular electronic state is not completely understood by the three-level fluctuation. The reactant may either lie in an ionic, closed-shell configuration or in a neutral radical state. The first scenario typically happens on metal surfaces, where the hydrogen-abstracted phthalocyanines or porphyrins accept charge from the substrate \cite{Sperl2011,Auwaerter2012, Kugel2017, Pham2016}. In contrast, on the \mos\ layer the electronic band gap may prevent such charge transfer. In the following, we show that we can distinguish between these possibilities by the spectroscopic signatures presented above and their comparison to simple DFT calculations of isolated molecules.

To justify the simple simulations of gas-phase molecules, we first simulate the electronic structure of \Pc\ and plot the resulting partial density of the states (PDOS) of the frontier molecular orbitals in \Figref{fig4}a. We label the highest occupied molecular state with H and the lowest unoccupied molecular state by L. We note that the LUMO+1 is quasi-degenerate (energy difference between LUMO and LUMO+1 amounts to $\sim$ 35\,meV). The HOMO-LUMO gap amounts to $\sim$ 2.2\,eV. The experimentally observed gap between the positive and negative ion resonance is slightly larger. Such a deviation from DFT calculations is expected, because the excitation requires the addition of an electron/hole. We also simulate  constant-height conductance images within the Bardeen approach using the DFT electronic structure (maps in \Figref{fig4}b).  The spatial distribution of the HOMO shows the eight-lobe structure as observed in experiment. The combined map of LUMO and LUMO+1 shows additional intensity in the center of \Pc, in agreement with the experimental conductance maps at the energy of the broad unoccupied state (see \Figref{fig1}d).

Next, we simulate the neutral, open-shell HPc molecule. Its PDOS exhibits an additional state ($h^*$) within the HOMO-LUMO gap of \Pc\ and the former H and L peaks are slightly shifted in energy (now labeled as $h$ and $l$, respectively).The LUMO is quasi-degenerate with the LUMO+1. Comparison to the experimental spectrum on the H-abstracted molecule shows that the most prominent change, namely the appearance of a new resonance $h^*$ within the former HOMO-LUMO gap, is captured by this model calculation (compare to \Figref{fig2}c). This suggests that the molecule lies in the neutral, open-shell configuration.
However, we find even more striking evidence for this identification from the fact that the conductance maps of $h$ and $h^*$ look identical as a result of imaging the SOMO and SUMO, respectively (simulated maps on the right side in \Figref{fig4}b and experimental maps in \Figref{fig2}c). In conclusion, the filled HOMO of \Pc\ converts to a singly-occupied molecular orbital (SOMO) upon hydrogen abstraction.
Moreover, we can associate the experimental energy difference between the SOMO and SUMO of $\sim$ 1.3\,eV  to the Coulomb repulsion energy of the electrons populating this state. Since the LUMO is lying outside of the energy gap of \mos\ and is therefore several hundred meV broad, we are not able to resolve any spin-splitting in the experiment (see spectra in \Figref{fig2}c).

At first sight, it may seem surprising that the radical state is imprinted in the delocalized HOMO instead of in a localized dangling $\sigma$ bond at the N site. However, it was recently shown that abstraction of both pyrrolic H atoms does not yield a bi-radical state at the N sites. Instead, the localized $\sigma$ states are filled by electrons from the HOMO \cite{Neel2016}. Similarly, in our case, the dangling bond is saturated by an electron from the HOMO leaving the HPc molecule in a delocalized radical state.
The stabilization of this radical state on the \mos\ layer is intriguing and stands in contrast to closed-shell states of H-abstracted phthalocyanines or porphyrins on metal surfaces. In those systems, the molecule--metal interactions lead to sizable structural relaxations of the molecule and charge transfer \cite{Sperl2011,Auwaerter2012, Kugel2017, Pham2016}. In contrast, the \mos\ layer suppresses structural deformations due to the absence of reactive bonding sites. Moreover, its band gap protects the radical state to be compensated by charge from the substrate.  More frequently than \mos, ionic layers (such as NaCl) or graphene have been employed for decoupling molecules from a metallic substrate. Interestingly, similar pulse experiments on phthalocyanine on graphene on Ir(111) did not lead to a single H abstraction, but to the simultaneous abstraction of both pyrrolic H atoms \cite{Neel2016}. Hence, the final state is again a closed-shell electronic configuration.

In conclusion, we have demonstrated that abstraction of a single H atom from the pyrrolic center of \Pc\ leads to the formation of a $\pi$-radical if the molecules lie on a decoupling layer of \mos\ on a metal substrate. The \mos\ takes on an important role in the stabilization of the radical state. When the molecules are in direct contact to a metal substrate, the molecules undergo significant structural relaxations such that the broken bond is compensated by a chemical bond to the substrate. The molecule thus looses its radical character. The passive S termination of \mos\ does not support bond formation. However, the dangling $\sigma$ bond is also not stable in this configuration. It is filled by intramolecular charge transfer from the HOMO, leaving behind a delocalized radical state.
Our study demonstrates the fully inert character of \mos\ towards otherwise highly reactive molecules. Hence, \mos\ may be used for studying intermolecular reactions where the substrate plays no significant role.

\section{Methods}

The Au(111) crystal was cleaned under ultra-high vacuum conditions by sputtering with Ne$^+$ ions at 1\,kV, followed by annealing to 800\,K. Monolayer-islands of \mos\ were grown \textit{in-situ} by deposition of Mo onto the atomically clean surface, followed by annealing the pre-covered sample at 820\,K under H$_2$S atmosphere \cite{Gronborg2015}. \Pc\ molecules were evaporated from a Knudsen cell heated to 680\,K onto the \mos -covered Au(111) sample held at a temperature below 120\,K.
The as-prepared sample was transferred into the STM, where all measurements were performed at a temperature of 4.6\,K. Differential-conductance spectra and maps were recorded with an open feedback-loop using lock-in detection with a modulation frequency of 721 Hz and a root mean square modulation amplitude between 1 and 10\,mV for the spectra (indicated in the caption for each spectrum) and 15\,mV for the maps.

The structure and electronic energy levels of isolated \Pc\ and \HPc\ molecules were simulated by density functional theory as implemented in the Gaussian 09 code using the b3pw91 functional and employing the 6-31++g(d,p) basis set \cite{Gaussian}. 
These structures served as the input for simulations of constant-height conductance maps by employing Bardeen`s approach using an s-wave tip with a work function of 5.3\,eV (simulating Au), and a tip molecule-distance of $\sim 4$\,\AA.
The PDOS was extracted from the calculated electronic structure using the code Multiwfn 3.3.9 \cite{multiwfn}.

\subsection{Supporting Information}
Further examples of H abstraction in \Pc\ molecules on \mos\ and switching properties of pyrrolic H atoms can be found in the Supporting Information, which is available free of charge on the ACS Publications website. 	
	
\section{Acknowledgments}
We gratefully acknowledge funding from the German Research Foundation within the collaborative research center TRR227 (project B05) and from the European Research Council for the Consolidator Grant "NanoSpin".

\bibliographystyle{apsrev4-1}
%

\clearpage

\setcounter{figure}{0}
\setcounter{section}{0}
\setcounter{equation}{0}
\setcounter{table}{0}
\renewcommand{\theequation}{S\arabic{equation}}
\renewcommand{\thefigure}{S\arabic{figure}}
	\renewcommand{\thetable}{S\arabic{table}}%
	\setcounter{section}{1}
	\renewcommand{\thesection}{S\arabic{section}}%

\onecolumngrid

\newcommand{\vsigma}{\mbox{\boldmath $\sigma$}}

\section*{\Large{Supporting Information}}

\section{Further examples of H abstraction in \Pc\ molecules on \mos}

\begin{figure}[h]
\begin{center}
\includegraphics[width=0.8\columnwidth]{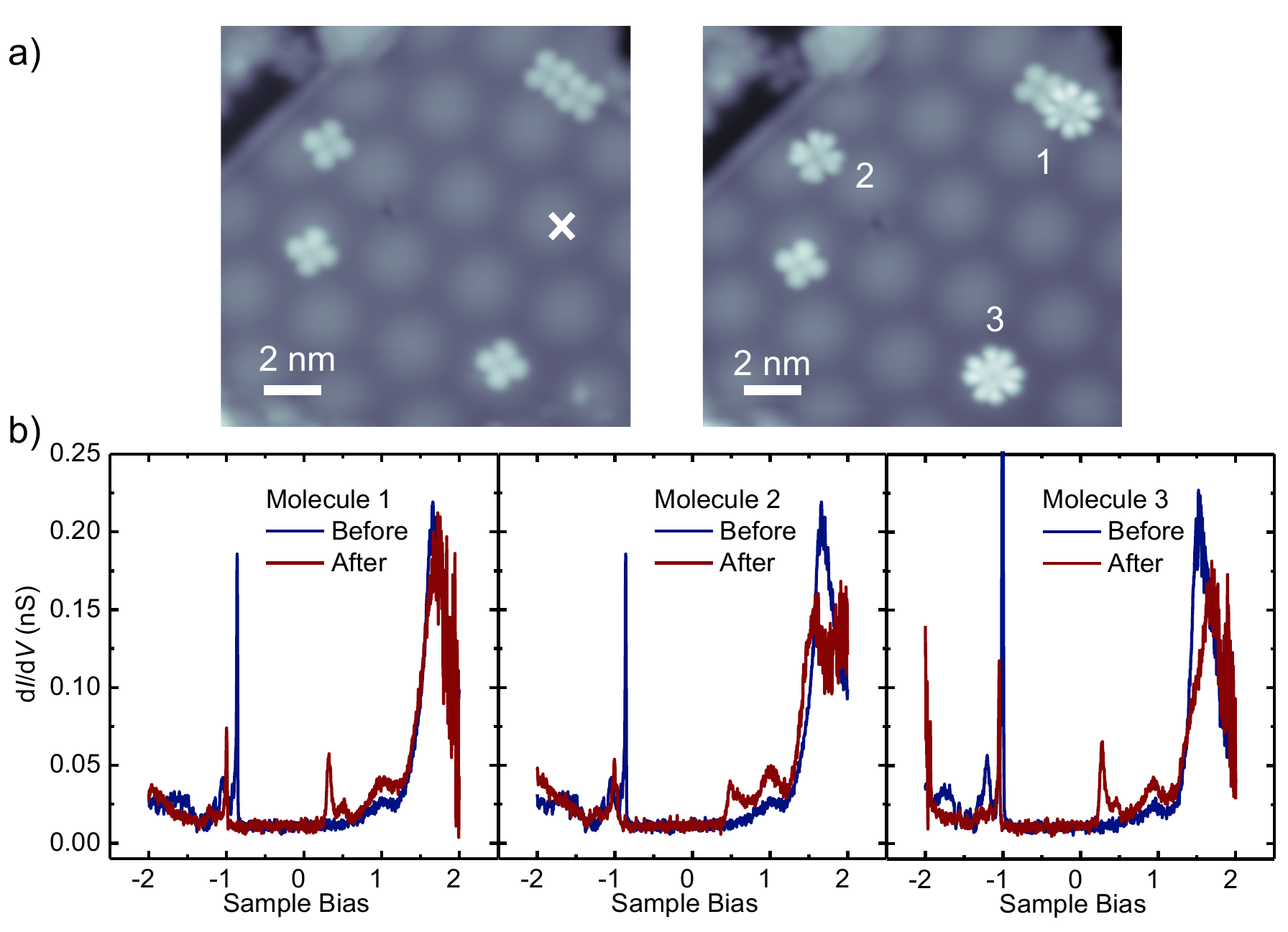}
\end{center}
\caption{a) STM topographies before and after a voltage pulse (3.5\,V, 1\,nA, 60\,s) was applied at the position marked by the white cross (set point: V\,=\,0.4\,V, I\,=\,20\,pA). b)  \didv spectra of the three molecules labeled by numbers in a), recorded before and after the voltage pulse.}
\label{figS1}
\end{figure}

As described in the main text, applying a voltage pulse in the vicinity of \Pc\ molecules on \mos\ may lead to H abstraction. \Figref{figS1} shows further examples of H abstraction after a voltage pulse of 3.5\,V, 1\,nA, has been launched for 60\,s at the position marked by the white cross. Scanning at a bias voltage of 0.4\,V reveals that three out of the five molecules in the scan range appear with the distinct eight-lobe structure reflecting the shape of the HOMO of the \Pc\ molecule. The observation of a considerable fraction of molecules being modified within a finite distance from the voltage pulse indicates a field-induced reaction mechanism \cite{SAlemani2006}.

The observation of the HOMO shape at positive bias voltage is indicative of a (partial) depopulation of the HOMO. As explained in more detail in the main text, we show that the modified molecules exhibit a single electron in the HOMO. 
\didv spectra recorded before and after the voltage pulse (\Figref{figS1}b) reflect the characteristic changes in the electronic structure that have also been described in the main text. Most importantly, all modified molecules exhibit a new resonance at positive bias voltage, with the precise value varying between 0.25\,V and 0.45\,V. The lower the energy of the corresponding state, the more narrow is its line shape, eventually even exhibiting vibronic resonances seen as higher-energy satellite peaks in \Figref{figS1}b. When the energy of this state approaches the broad edge of the conduction band, the peak width increases due to faster relaxation rates of the excited electronic state. We ascribe the variation of the energy level alignment to the differences in adsorption site on the moir$\acute{\mathrm{e}}$ structure and/or influence of neighboring molecules (see e.g. molecule 1). Similar to the variations of the SUMO state, we also find variations in energy of the SOMO state, corroborating this interpretation.

\section{Switching properties of pyrrolic H atoms}  

\begin{figure}[h]
\begin{center}
\includegraphics[width=0.8\columnwidth]{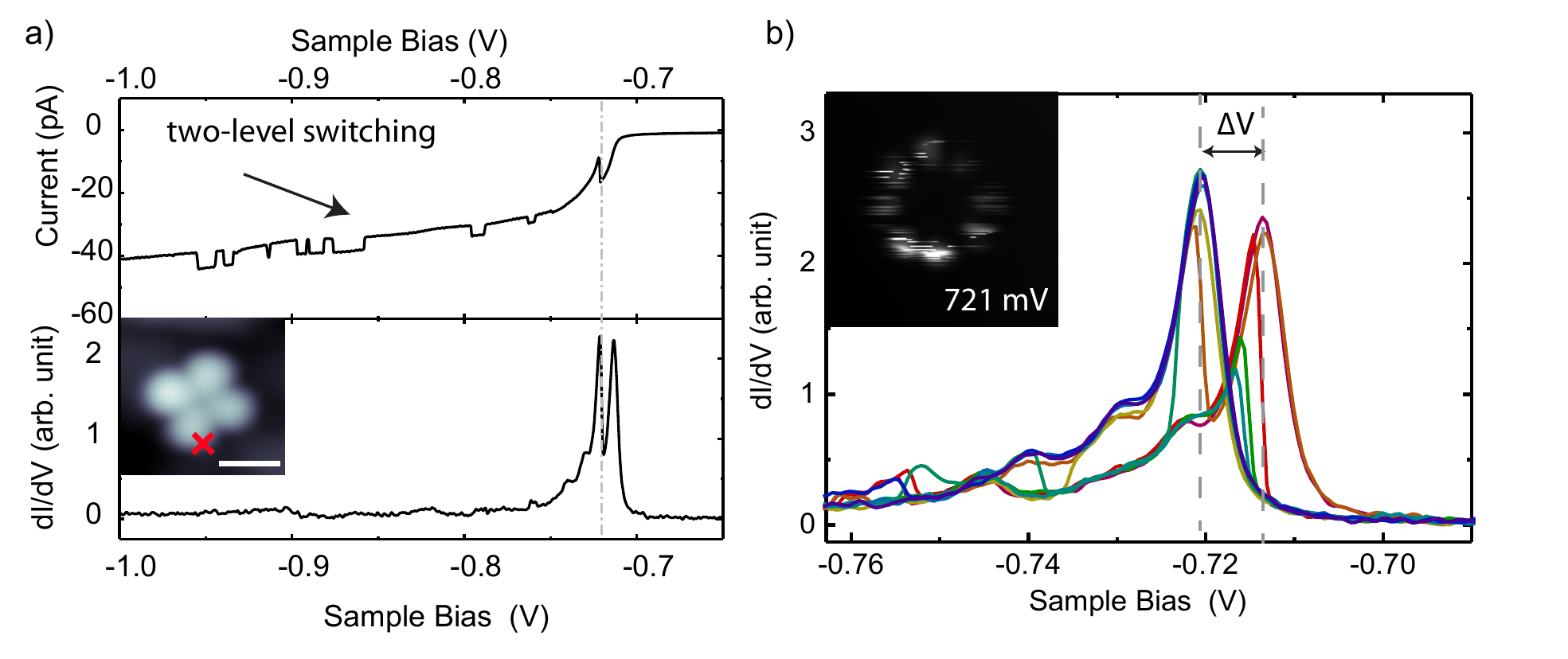}
\end{center}
\caption{a) I(V) curve and simultaneously recorded \didv curve on a \Pc\ molecule shown as inset (0.4\,V, 10\,pA, scale bar is 1nm), at the position marked by the red cross. The tautomerization reaction is seen as a two-level fluctuation in the current trace. A pronounced change in the \didv curve is detected at the energy of the HOMO indicated by the gray dashed line (set point: 1\,V,  40\,pA). b) Representation of 10 successive \didv spectra recorded over the same molecule and with the tip at the same position (set point: 1\,V,  40\,pA, $V_\text{mod}$ = 1\, mV). The inset shows a \didv map recorded on the same molecule at the energy of one the peak resolved in the spectra ($V_\text{mod}$=3\,mV).}
\label{figS2}
\end{figure}

As explained in more detail in the main manuscript, we observe two-level current fluctuations on as-deposited \Pc\ molecules whereas three-level fluctuations occur on the modified molecules. The occurrence of three-level fluctuations has been ascribed to H abstraction in the pyrrole core. The switching properties of the H atoms in the pyrrole have been intensively studied both for the \Pc\ and HPc \cite{SKugel2016,SKugel2017} and other similar macrocycle molecules on metal substrates. In this section, we emphasize some differences in the switching properties on \mos\ as compared to a metal substrate. 

On a metal substrate, the switching of the inner hydrogen atoms is accompanied by a geometrical reorganization of the molecule due to the interaction with the metal \cite{SSperl2011}. Thus, when imaged at low bias voltages, two (one) lobes of the \Pc\ (HPc) molecule appear higher, indicating the position of the inner hydrogen. The 'reading' of the tautomerization state is mainly a topographic effect. On \mos, the situation is different: at low bias voltages (inside the gap of the molecule) the molecule (\Pc\ or HPc) appears with a 4-fold symmetry (see Fig.1 and 2 in the main manuscript). This is explained by a flat adsorption of the molecule on \mos\ (similar to graphene \cite{SNeel2016}), which is possible when the interaction with the substrate is very small. Therefore, in the absence of structural relaxations, the tautomerization state cannot be 'read' as a topographic effect. 

Instead, we can only observe the electronic effect of the tautomerization of the molecules on \mos. We illustrate this effect at the energy of the HOMO of \Pc. \Figref{figS2}a shows a typical I(V) curve and the simultaneously recorded \didv spectrum of \Pc. The I(V) curve reflects a two-level switching associated to the tautomerization reaction. The variations in the \didv spectrum become very pronounced when the bias voltage reaches the HOMO energy (gray dashed line in \Figref{figS2}a). To understand the large variations in the conductance at this energy range (as compared to higher bias voltages), we recorded ten  successive spectra on the same molecule (\Figref{figS2}b). This set of spectra reflects two distinct peaks, which are separated by $\Delta V \sim $ 7\,mV. Moreover, the peaks are of different height. Hence, the two peaks reveal a shift in the HOMO energy when the H atoms are placed in different locations of the molecule. Such a behavior can only be explained by a reduced symmetry of the molecule on the substrate including a different interaction of the HOMO with the substrate in the two orientations. We ascribe this to the different effective orientations of the HOMO with respect to the underlying atomic lattice and moir$\acute{\mathrm{e}}$ structure of \mos.
Furthermore, the change of intensity is a result of the twofold symmetry of the HOMO state with the symmetry axis given by the position of the two H atoms, and therefore, depends on tip position. 

While recording conductance maps of this state, the modification of the electronic structure is also observable. In the inset of \Figref{figS2}b, we show one of these maps recorded at the energy of one of the peaks with a small lock-in modulation amplitude of 3\,mV. In this map we observe a large telegraph noise due to the shift of the HOMO peak with the tautomerization. For the conductance maps presented in the manuscript (Figure 1 and 2), we used a larger modulation amplitude (15\,mV) in order to reduce this effect and reveal more clearly the eight-lobed shape of the orbital. 

\begin{figure}[h]
\begin{center}
\includegraphics[width=0.8\columnwidth]{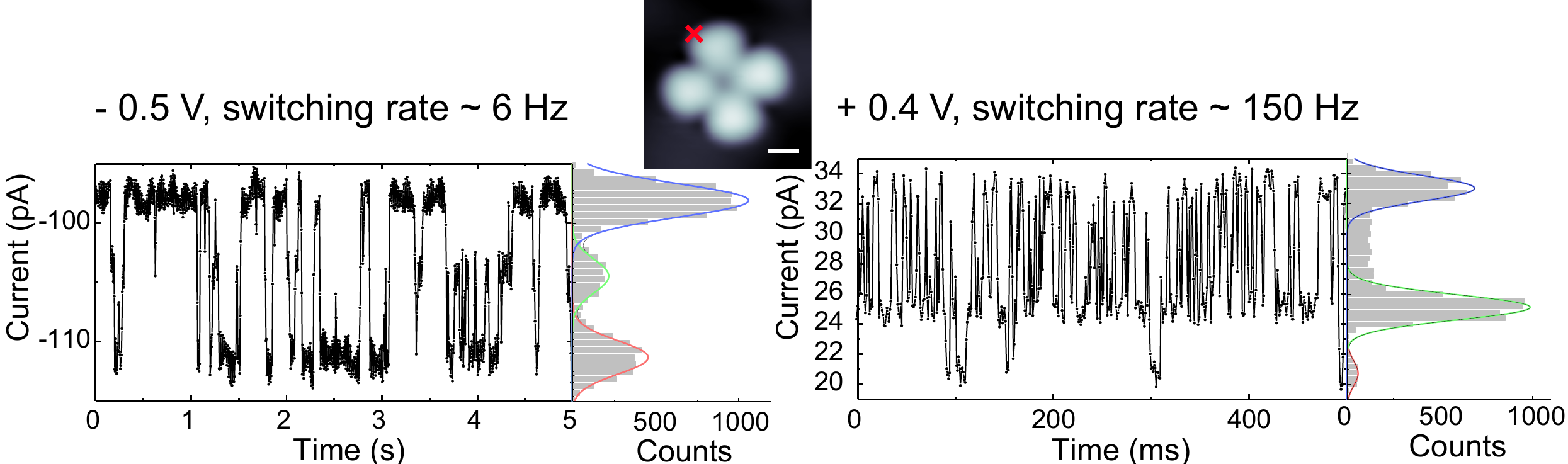}
\end{center}
\caption{Current traces and associated histograms recorded on the same HPc molecule (image as inset, 0.1\,V, 30\,pA, scale bar 0.5\,nm) and with the tip parked on the same position (red cross) at different applied voltages: -0.5\,V, initial set point 100\,pA (left) and +0.4\,V, initial set point 30\,pA (right). For readability, the traces show only a small sample of the complete measurements, which are represented in the histograms (10000 points for a measurements time of 50 and 10\,s respectively). Each histogram is fitted with three Gaussian functions.}
\label{figS3}
\end{figure}
     
Next, we discuss the switching efficiency. For \Pc\ and HPc adsorbed on metal substrates, it has been shown that the switching rate is strongly increased for HPc compared to \Pc\ (by almost three order of magnitude) \cite{SKugel2016,SKugel2017}. This is explained by the reduction of the potential barrier for a single hydrogen switching compared to the collective switching of two hydrogen atoms. Moreover, it has been shown that the voltage dependance of the switching is symmetric in bias voltage (thresholds of increasing rates at $\pm$0.1, $\pm$0.4 and $\pm$0.9\,V). This behavior has been explained by a vibration-mediated switching mechanism.

For \Pc\ on \mos, the switching rates behave differently. While we do not observe any switching at voltages within the HOMO-LUMO gap (at maximum 1\,nA), the switching rates increase drastically at the onset of molecular resonances (at -0.8 or +1.4\,V) and are then higher than on a metal (\textit{e.g.}, at -1.1\,V: 2\,Hz at  50\,pA on \mos, $vs.$ 5\,Hz at 1\,nA on Ag\cite{SKugel2017}). Tautomerization events occur even at low currents above the resonance (see I(V) curve in \Figref{figS2}). From the threshold behavior we can exclude a vibration-mediated mechanism. A similar conclusion was drawn for the case of naphthalocyanine molecules on NaCl \cite{SLiljeroth2007}. 

The switching behavior of HPc is again somewhat different. In contrast to \Pc, we observe switching within the HOMO-LUMO gap (see, \textit{e.g.}, current trace at -0.5\,V in \Figref{figS3} and 0.2\,V in Figure 3 of the main manuscript). When increasing the bias voltage, the switching rate is strongly enhanced (from few Hz at 0.2\,V to few hundred Hz at 0.4\,V). Moreover, there is a strong bias asymmetry (few Hz at -0.5\,V \textit{vs.} few hundred Hz at 0.4\,V). Similarly to the measurements in Figure 3 of the manuscript, we observe a three-level switching. For the current trace at 0.4\,V, this observation is not so clear without the histogram of the current distribution, because of the fast switching rate ($\approx$ 150\,Hz), which is close to the limit of the bandwidth of the I(V) converter (1kHz).

\providecommand{\latin}[1]{#1}
\makeatletter
\providecommand{\doi}
  {\begingroup\let\do\@makeother\dospecials
  \catcode`\{=1 \catcode`\}=2 \doi@aux}
\providecommand{\doi@aux}[1]{\endgroup\texttt{#1}}
\makeatother
\providecommand*\mcitethebibliography{\thebibliography}
\csname @ifundefined\endcsname{endmcitethebibliography}
  {\let\endmcitethebibliography\endthebibliography}{}

\end{document}